\documentclass[prb,twocolumn,eqsecnum,showpacs,amsmath,amssymb,aps]{revtex4}

\usepackage{graphicx}
\usepackage{dcolumn}
\usepackage{bm}
\usepackage{epsfig}

\begin{document}


\title{Even-odd parity effects in conductance and shot noise 
of metal-atomic wire-metal(superconducting) junctions}

\author{Tae-Suk Kim$^{1,2,3}$ and S. Hershfield$^{3}$}
\affiliation{$^{1}$ Institute of Physics and Applied Physics, Yonsei University, 
  Seoul 120-749, Korea \\
 $^{2}$ School of Physics, Korea Institute for Advanced Study, 
Seoul 130-012, Korea \\
 $^{3}$ Department of Physics, University of Florida, Gainesville FL 32611-8440 }

\date{\today}

\begin{abstract}
In this paper, we study the conductance and shot noise 
in transport through 
a multi-site system in a two terminal configuration. The dependence of 
the transport on the number of atoms in the atomic wire is investigated 
using a tight-binding Hamiltonian and the nonequilibrium Green's function
method. 
In addition to reproducing the even-odd behavior in the transmission probability 
at the Fermi energy or the linear response conductance
in the normal-atomic wire-normal metallic(NAN) junctions, 
we find the following: (i) The shot noise is larger in the even-numbered 
atomic wire than in the odd-numbered wire. 
(ii) The Andreev conductance displays the same even-odd parity effects 
in the normal-atomic wire-superconducting(NAS) junctions. 
In general, the conductance is higher in the odd-numbered atomic wire
than in the even-numbered wire. 
When the number of sites ($N$) is odd and the atomic wire is mirror symmetric
with respect to the center of the atomic wire, the conductance does not 
depend on the details of the hopping matrices in the atomic wire, but 
is solely determined by the coupling strength to the two leads. 
When $N$ is even, the conductance is sensitive to the values of the hopping 
matrices.  
\end{abstract}

\pacs{73.63.Nm, 74.50.+r}

\maketitle

\section{Introduction}
 Electron transport through an atomic wire is realized in scanning tunneling
microscope(STM) experiments\cite{n783,n780} and break junctions\cite{break}.
An array of quantum dots\cite{qdots} can be considered as an artificial 
atomic wire. 
In the STM experiments, a true atomic wire was realized as  
the dipped gold STM tip is slowly retracted from the Au substrate. 
Just before breaking, the metallic contacts become thinner down to one atom wide
and consisting of a few bridging atoms. 
Yanson and others\cite{n783} succeeded in forming chains of gold atoms
up to seven atoms long. 
The observed conductance\cite{n783,n780,spain1}
stayed close to one unit of the conductance quantum
before breaking.

 A quantum dot is a collection of electrons within a small space
and can be considered as an artificial atom.
Since electrons are confined to the small region, 
electrons experience the strong Coulomb repulsion 
and their energy level spectrum is discrete.
Like electrons in atoms, the charge and spin of electrons are 
quantized in quantum dots. 
In contrast to the atoms in the periodic table, 
it is easy to control the energy level spacing, 
the strength of Coulomb interaction, 
the number of electrons (charge), spin, etc in quantum dots. 
When quantum dots are arranged as a linear chain, the system can be 
considered as an ``atomic" wire. 
Some I-V curves have been measured in an array of quantum dots. 
The splitting of the Coulomb peaks\cite{qdots}
 was observed in the conductance when the interdot tunneling
rate is strong compared with the coupling to external electrodes.

\begin{figure}[h]
\protect\centerline{\epsfxsize=0.40\textwidth \epsfbox{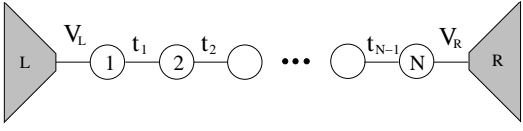}}
\caption{Schematic display of a serial multi-site system.
\label{smdots}}
\end{figure}

 In this paper, we study the conductance and shot noise
in transport through an atomic wire in a two termial configuration.  
The atomic wire in our study consists of 
the monovalent atoms like Na, Au, Ag, Cu, etc. 
and is modeled by a single orbital tight-binding Hamiltonian(see Fig.~\ref{smdots}.)
When two metallic leads, connected to the atomic wire, are
both normal, we study the dependence of linear response conductance 
and shot noise on the number of atoms in the atomic chain. 
When one lead is normal and the other is superconducting, 
the dependence of the Andreev conductance on the number of 
atoms in the chain is studied. 
Our discussion is confined to the linear response regime or 
to the case of weak source-drain bias voltage.

 The even-odd behavior of the linear response conductance or 
oscillation of the conductance as a function of the number of atoms
in the atomic wire was theoretically studied using {\it ab initio}
calculations\cite{lang,kaist}.
In heterogeneous carbon atomic wires\cite{lang} where
the carbon atomic wires
are connected to two metal electrodes, the linear response
conductance is found to be oscillating as a function of the number of 
C atoms in the atomic wire. The conductance is smaller than the conductance
quantum $G_Q = 2e^2/h$ and is higher in the odd-numbered atomic wires than 
in the even-numbered systems. 
On the other hand, the conductance is very close to $G_Q$ 
in the homogeneous sodium atomic wire\cite{kaist} when the number of atoms in 
the wire is odd.  
In general, the conductance is higher for the atomic wire with an odd number
of atoms than with an even number of atoms. 
This even-odd parity effect in the 
conductance can be understood based upon the tight-binding Hamiltonian
\cite{tight,condmat} 
for the atomic wire. 
In addition to reproducing the even-odd behavior
of the conductance in the metal-atomic wire-metal junction, we 
also find the same oscillation in the Andreev conductance as a function of 
the number of atoms in the atomic wire for the metal-atomic wire-superconducting
junctions.  
When the number of sites ($N$) is odd and the atomic wire is mirror symmetric
with respect to the center of the atomic wire, the conductance does not 
depend on the details of the hopping matrices in the atomic wire, but 
is solely determined by the coupling strength to the two leads. 
When $N$ is even, the conductance is sensitive to the values of the hopping 
matrices.

In recent STM experiments of gold nanocontacts\cite{spain1}, a small 
variation of conductance within the region of a unit conductance quantum
was observed (Fig.~1a of reference\cite{spain1})
 as the STM tip is stretched before breaking.
Since the conductance is close to the unit conductance quantum, 
the gold chain is one atom thick. A small variation but abrupt jump in the 
conductance might originate from an addition of an extra gold atom to the chain.
That is, the observed conductance variation may be a signature of 
the even-odd parity effects in the atomic wire.

 This paper is organized as follows.
In Sec.~\ref{NN_sec}, we study the transport through multi-sites 
which are connected to the two normal metallic leads.
We consider the normal-superconducting junctions passing through the 
multi-sites in Sec.~\ref{NS_sec}.  
Two sections are the separate works and can be read independently. 
Our study is summarized in Sec.~\ref{sum_sec}.

\section{\label{NN_sec} Normal-normal metallic(N-N) junctions connected by multi-sites}
 In this section we study 
the current and its noise in the transport through multi-resonant sites
in a two-terminal configuration where two leads are in the normal metallic state.
The formalism is most conveniently simplified using the matrix notations. 
The current is given in a well-known Landauer-B\"{u}ttiker form
\cite{landauer,buttiker}
and the shot noise is also expressed in the familiar binomial form\cite{shotnoise}
except for some special cases.

 Applying the formalism to the serial multi-site system or the atomic wire
connected to two metallic leads, we give a simple explanation 
of the even-odd behavior in the transmission probability at the Fermi energy 
or linear response conductance which was reported 
in the recent theoretical works\cite{lang,kaist}.  
In a mirror symmetric atomic wire, the transmission probability 
at the Fermi energy takes the same value which depends only on the 
coupling strength to the two leads,  when the number of sites is odd. 
On the other hand, for the system with an even number of sites, 
the transmission probability 
depends on the detailed values of the hopping matrices between neighboring 
sites. This even-odd behavior can be understood in terms of the level-splitting 
in quantum physics and the particle-hole symmetry in the energy spectrum 
when all the energy levels of atoms lie at the Fermi energy. 
We also find that the shot noise in the transport 
through multi-site system is oscillating as a function of 
the number of atoms in the atomic wire.

 The multi-site system made of monovalent atoms (Na, Au, Ag, Cu) can be described by 
the model Hamiltonian, 
\begin{eqnarray}
H_{cb} &=& \sum_{p=L,R} \sum_{k} \epsilon_{pk} c_{pk}^{\dag} c_{pk}, \\
H_d &=& \sum_{i=1}^{N} E_i d_i^{\dag} d_i + \sum_{i\neq j} W_{ij} d_i^{\dag} d_j, \\
H_1 &=& \sum_{pk}\sum_{i} \left[ V_{pi} c_{pk}^{\dag} d_i 
      + V_{ip} d_{i}^{\dag} c_{pk} \right].
\end{eqnarray}
Since the electron spin does not play any role 
in the N-N junctions, the spin index will be suppressed in this section. 
Electrons can flow from one lead to the other one through all the sites when 
a finite source-drain voltage is applied. All the sites are connected 
to each other via tunneling. 
The serial multi-site model is recovered when 
$W_{ij} = 0$ unless $j=i\pm 1$ and $V_{iL} = V_L \delta_{i,1}$ and 
$V_{iR} = V_R \delta_{i,N}$. 
The algebra is most conveniently simplified and becomes compact 
when the Hamiltonian is written in a matrix form by introducing 
\begin{eqnarray}
\Psi_d^{\dag} 
 &=& \begin{pmatrix} d_1^{\dag} & d_2^{\dag} & \cdots & d_N^{\dag} \end{pmatrix}, \\
V_p^{\dag} 
 &=& \begin{pmatrix} V_{p1} & V_{p2} & \cdots & V_{pN} \end{pmatrix}, 
\end{eqnarray}
and $H_{d,ii} = E_i, H_{d,ij} = W_{ij}$. With these notations, the model Hamiltonian
becomes
\begin{eqnarray}
H_d &=& \Psi_d^{\dag} H_d \Phi_d, \\
H_1 &=& \sum_{pk} \left[ c_{pk}^{\dag} V_p^{\dag} \Psi_d^{\dag} 
           + \Psi_d^{\dag} V_p c_{pk} \right]. 
\end{eqnarray}

\subsection{Current and shot noise}
The current operator which is defined as the variation per unit time  
 of electron charges in the left or right lead can be written 
in a matrix form as
\begin{eqnarray}
\hat{I}_p 
 &=& \frac{e}{i\hbar} [ \hat{N}_p, H ]
 = \frac{e}{i\hbar} \sum_{k} \left[ c_{pk}^{\dag} V_p^{\dag} \Psi_d 
    - \Psi_d^{\dag} V_p c_{pk} \right]. 
\end{eqnarray}
After thermal averaging the current operator, current is written 
in terms of the mixed Green's function out of equilibrium.
\begin{eqnarray}
I_p &=& 2e \sum_{k} \mbox{Im} \mbox{Tr} G_{dp}^{<} (t, kt) V_{p}^{\dag}, \\
i\hbar G_{dp} (t, kt') 
 &=& \langle T_c \Psi_d(t) c_{pk}^{\dag} (t') \rangle. 
\end{eqnarray}
Here $T_c$ means contour-ordering where the contour encloses 
the real-time axis\cite{keldysh,langreth}. 
Using the Dyson equation for the mixed Green's function $G_{dp}$
and the Green's functions of two leads, 
the above expression of current is 
reduced to a simple compact Landauer-B\"{u}ttiker
form\cite{landauer,buttiker}.
\begin{eqnarray}
I_L &=& \frac{2e}{h} \int d\epsilon ~T(\epsilon) [ f_R(\epsilon) - f_L(\epsilon) ], \\
\label{tspec}
T(\epsilon) 
 &=& 4 \mbox{Re} \mbox{Tr} [ \Gamma_L D^r(\epsilon) \Gamma_R D^a(\epsilon) ].
\end{eqnarray}
The factor $2$ is recovered in $I_L$ to take into account two directions
of an electron spin. $\Gamma_p = \pi N_p V_pV_p^{\dag} (p=L,R)$
is the linewidth matrix. $D^{r,a}$ is the retarded (advanced) part of 
Green's function of the multi-site system which is defined by
\begin{eqnarray}
i\hbar D(t,t') &=& \langle T_c \Psi_d(t) \Psi_d^{\dag}(t') \rangle.
\end{eqnarray}
$f_p(p=L,R)$ is the Fermi-Dirac thermal distribution function in the 
left and right leads.

 The current noise is defined as the current-current correlation function.
\begin{eqnarray}
S(t,t') &=& \langle \delta \hat{I}(t) \delta \hat{I}(t') \rangle 
  + \langle \delta \hat{I}(t') \delta \hat{I}(t) \rangle. 
\end{eqnarray}
Here $\delta \hat{I}(t) = \hat{I}(t) - \langle \hat{I}(t) \rangle$ is 
the current fluctuation operator.  To compute the current noise, 
we introduce the current Green's function,
\begin{eqnarray}
i\hbar G_{II} (t,t') 
 &=& \langle T_c \delta \hat{I}(t) \delta \hat{I}(t') \rangle. 
\end{eqnarray}
Using the greater and lesser current Green's functions, 
$G_{II}^{>} (t,t') = \langle \delta \hat{I}(t) \delta \hat{I}(t') \rangle$ 
and $G_{II}^{<} (t,t') = -\langle \delta \hat{I}(t') \delta \hat{I}(t) \rangle$, 
the current noise can be written as
\begin{eqnarray}
S(t,t') &=& \hbar G_{II}^{>} (t,t') - \hbar G_{II}^{<}(t',t).
\end{eqnarray}
Since $G_{II}^{>}(t,t') = - G_{II}^{<} (t',t)$,
calculation of only the lesser current Green's function
is needed to find the current noise.
For the noninteracting electrons in the multi-site system, 
there are only six Feynman diagrams, as shown in Fig.~\ref{noise}, 
contributing to the current noise. Detailed calculations of each diagram
are described in the Appendix~\ref{app_noise}. 
The current noise expression is given by the equation,  
\begin{eqnarray}
\label{shoteqn}
S_0 &=& \frac{4e^2}{h} \int d\epsilon [f_L\bar{f}_L + f_R\bar{f}_R]
   \mbox{Tr} {\bf T}^2 \nonumber\\
 && + \frac{4e^2}{h} \int d\epsilon [f_L\bar{f}_R + \bar{f}_L f_R]
   \left[ \mbox{Tr} {\bf T} - \mbox{Tr} {\bf T}^2 \right].
\end{eqnarray}
Additional factor $2$ is multiplied to get the spinful result of $S_0$. 
Here ${\bf T} = 4 \Gamma_L D^r \Gamma_R D^a$ and 
$f_p(p=L,R)$ is the Fermi-Dirac thermal function in the left and right
electrodes and $\bar{f}_p = 1 - f_p$.
As expected, the current noise $S_0$ is symmetrical in the lead indices
$L$ and $R$. 
In the expression of current, the transmission probability $T(\epsilon)$ is
defined as
$T(\epsilon) = \mbox{Tr} {\bf T}$.
Manipulating the thermal functions, we can rewrite the expression 
of $S_0$ as
\begin{eqnarray}
\label{shotnoise_1}
S_0 &=& \frac{4e^2}{h} \int d\epsilon [f_L\bar{f}_L + f_R\bar{f}_R] T(\epsilon)
   \nonumber\\
 && + \frac{4e^2}{h} \int d\epsilon [f_L - f_R]^2 \mbox{Tr} 
   \left\{ {\bf T} (\epsilon) - [{\bf T}(\epsilon) ]^2 \right\}. 
\end{eqnarray}
The first line is the thermal or Johnson noise and remains finite 
at nonzero temperature. 
The second line is nonzero only out of equilibrium and is the so-called 
shot noise deriving from the discreteness of electron's charge. 
We note that the integrand in the second line is simplified 
to the familiar binomial expression\cite{shotnoise}
 $T(\epsilon) [ 1 - T(\epsilon)]$
except when both $\mbox{det}\Gamma_L \neq 0$ and $\mbox{det}\Gamma_R \neq 0$. 
This point can be easily proved by noting that ${\bf T}$ and ${\bf T}^2$
can be diagonalized at the same time and the trace is a sum of 
diagonal elements. 
Only one diagonal element is non-zero after diagonalization
and all the others are zero.  
As concrete examples, the familiar binomial expression of $S_0$ 
is obtained for the two-site or the serial multi-site systems.  
Note that for a $2\times 2$ matrix $A$ we have the identity
\begin{eqnarray}
\mbox{Tr} A^2 &=& [\mbox{Tr} A]^2 - 2\mbox{det} A.
\end{eqnarray}
However, $\mbox{det} {\bf T} = 0$ since $\mbox{det} {\bf \Gamma}_L = 0 
=\mbox{det} {\bf \Gamma}_R$. That is, the current noise for the two-site
system is reduced to the familiar binomial expression.
There are exceptional cases\cite{utah} of two-site system for which 
$\mbox{det} {\bf \Gamma}_L \neq 0 \neq \mbox{det} {\bf \Gamma}_R$.
In this case, the familiar binomial expression of the shot noise 
is not recovered.

\subsection{Atomic wire: serial multi-sites}
 In this section, we apply the formalism of the 
previous section to the metal-atomic wire-metal junctions 
for which the atomic wire is modeled by the tight-binding Hamiltonian. 
In a serial configuration of multi-sites,
the hopping matrices are non-zero only between two neighboring sites,
\begin{eqnarray}
W_{ij} &=& t_i \delta_{i+1,j} + t_j^* \delta_{i,j+1}.
\end{eqnarray}
Since only the left (right)-most site is connected to the left (right) lead, 
the linewidth matrices are simplified as
\begin{eqnarray}
\Gamma_{L,ij} &=& \Gamma_L ~\delta_{i,1}\delta_{j,1}, \\
\Gamma_{R,ij} &=& \Gamma_R ~\delta_{i,N}\delta_{j,N}.
\end{eqnarray}
Under this condition, the transmission probability is given by the expression,
\begin{eqnarray}
T(\epsilon) 
 &=& 4 \mbox{Tr} \Gamma_L D^{r} \Gamma_R D^a 
 = 4\Gamma_L \Gamma_R ~ | D_{1N}^{r} (\epsilon)|^2.
\end{eqnarray}
Here $D_{1N}^{r}$ is the retarded part of the Green's function defined by 
\begin{eqnarray}
i\hbar D_{1N}(t,t') 
 &=& \langle T_c d_1(t) d_N^{\dag}(t') \rangle, 
\end{eqnarray}
and is given by the $(1,N)$ element of 
$D^r(\epsilon) = [\epsilon - H_d  + i\Gamma ]^{-1} $. 
Here $\Gamma = \Gamma_L + \Gamma_R$ is the total linewidth matrix.
Using the cofactor of the matrix, we find $D_{1N}^{r}$,
\begin{eqnarray}
D_{1N}^{r} (\epsilon) 
 &=& [\epsilon - H_d + i \Gamma]^{-1}_{1N} \nonumber\\
 &=& \frac{t_1 t_2 \cdots t_{N-1}}
           {Z_N(\epsilon)}, \\
\label{det}
Z_N(\epsilon) 
 &=& \mbox{det} [ \epsilon - H_d + i \Gamma].
\end{eqnarray}
The determinant $Z_N$ is calculated iteratively.

 Consider the case of $E_i = 0$ for all $i=1,2,\cdots,N$ and calculate 
the transmission probability at the Fermi energy. 
This particle-hole symmetric energy structure seems to be realized 
in the {\it ab initio} calculation of the homogeneous sodium 
atomic wire\cite{kaist}. 
Depending on the parity of the number of sites, the determinant 
takes on different forms, 
\begin{eqnarray}
Z_{2N} (0)
 &=& (-1)^N \Gamma_L \Gamma_R |t_2 t_4 \cdots t_{2N-2}|^2 \nonumber\\
 && + (-1)^N |t_1 t_3 \cdots t_{2N-1}|^2, \\
Z_{2N+1} (0)
 &=& i(-1)^N \Gamma_R |t_1t_3 \cdots t_{2N-1}|^2 \nonumber\\
 &&   + i(-1)^N \Gamma_L |t_2 t_4 \cdots t_{2N}|^2.
\end{eqnarray}
Accordingly, the transmission probability has the even-odd parity 
dependence:
\begin{eqnarray}
T_{2N}(0) 
 &=& \frac{4\Gamma_L \Gamma_R |t_1t_2\cdots t_{2N-1}|^2}
    { [ \Gamma_L\Gamma_R|t_2t_4\cdots t_{2N-2}|^2 
      + |t_1t_3\cdots t_{2N-1}|^2 ]^2 }, \\
T_2(0) &=& \frac{4\Gamma_L\Gamma_R |t_1|^2}
            {[\Gamma_L\Gamma_R + |t_1|^2]^2},
\end{eqnarray}
and 
\begin{eqnarray}
T_{2N+1}(0) 
 &=& \frac{4\Gamma_L \Gamma_R |t_1t_2\cdots t_{2N}|^2}
    { [ \Gamma_L |t_2t_4\cdots t_{2N}|^2 
      + \Gamma_R |t_1t_3\cdots t_{2N-1}|^2 ]^2 }, \\
T_1(0) &=& \frac{4\Gamma_L\Gamma_R}
            {[\Gamma_L + \Gamma_R]^2}.
\end{eqnarray}
The conditions of perfect transmission can be readily found from the above 
equations.

When the hopping matrices are mirror symmetric with respect to the 
the center of the atomic wire, the above expression of $T_N(0)$ for odd $N$ is 
further simplified.
\begin{eqnarray}
T_{2N+1}(0) &=& \frac{4\Gamma_L\Gamma_R} {[\Gamma_L + \Gamma_R]^2}. 
\end{eqnarray}
Note that this value does not depend on the detailed values of 
the hopping matrices between neighboring sites. 
On the other hand, the transmission probability $T_N(0)$ for even $N$
depends on the detailed values of hopping matrices.

When the hopping matrices between neighboring sites are all equal,
$t_i = t$ for all $i$'s, 
the transmission probability $T_N(0)$ is further simplified to
\begin{eqnarray}
T_{2N}(0) 
 &=& \frac{4\Gamma_L \Gamma_R t^2} {[t^2 + \Gamma_L \Gamma_R]^2}, \\
T_{2N+1}(0) 
 &=& \frac{4\Gamma_L\Gamma_R} {[\Gamma_L + \Gamma_R]^2}.
\end{eqnarray}
The transmission probability at the Fermi energy in the even-numbered atomic wire
still depends on the hopping matrix.
Perfect transmission is achieved when $\Gamma_L = \Gamma_R$ in the 
odd-numbered atomic wire, and when $t^2 = \Gamma_L \Gamma_R$ in the 
even-numbered atomic wire.

 The even-odd parity behavior of the transmission probability 
at the Fermi energy is a consequence of the energy level splitting 
in quantum physics and the particle-hole symmetry of atomic energy levels. 
In the particle-hole symmetric case, all the atomic energy levels
lie at the Fermi energy ($E_i=0$) when the hopping between atoms
and the coupling to the leads are turned off. 
When hopping between neighboring atoms in the atomic wire is turned on, 
the degeneracy in atomic energy levels is lifted and the energy level
structure still keeps the particle-hole symmetry
(see the Appendix~\ref{app_det}.) 
Note that the values of $t_i$'s have no influence on the particle-hole 
symmetry of split atomic energy levels.

 In the odd-numbered atomic wire, one level always lies at the Fermi energy 
and all others are distributed symmetrically with respect to the Fermi energy($E_F$).
Note that $Z_{2N+1}(0)$ is equal to $0$ when $\Gamma_{L,R} = 0$. 
This vanishing determinant means that at least one energy level lies at $E_F$.
Since no degeneracy remains in the presence of hopping between atomic sites, 
only one level is located at the Fermi energy.
When the coupling to external leads is turned on, the discrete energy 
levels become broadened. 
The transmission probability $T_N(\epsilon)$ for odd $N$ is always 
peaked at $\epsilon = 0$ or the Fermi energy.

 On the other hand, all energy levels are distributed symmetrically 
with respect to the Fermi energy without any level at $E_F$  
for the even-numbered atomic wires. 
Since $Z_{2N}(0)$ is not equal to $0$ when $\Gamma_{L,R} = 0$, 
the degeneracy-lifted atomic energy levels cannot lie at the Fermi energy
in the even-numbered atomic wire.
Though no energy level is present at the Fermi energy for the even-numbered 
atomic wire, either a dip or a peak in $T_N(\epsilon)$ for even $N$ 
is possible at the Fermi energy $\epsilon=0$ 
when $\Gamma_{L,R} \neq 0$, depending on the relative
magnitudes of the hopping integral (level-splitting) 
in the atomic wire and the coupling strength (linewidth)
to the leads. When $t_i = t$ for all $i$'s and $\Gamma_L = \Gamma_R$, 
the transmission probability is dipped(peaked) at the Fermi energy($\epsilon=0$)  
when $t >(\leq) \Gamma_{L,R}$, respectively.

 To illustrate the even-odd parity dependence of the transmission probability, 
$T_N(\epsilon)$ is displayed in Fig.~\ref{nntprob} 
for differing $N$'s (the number of atoms in the atomic wire). 
To reproduce the results of the {\it ab initio} calculations\cite{lang,kaist},
model parameters are chosen such that 
$E_i = 0$ for all sites and all tight-binding hopping matrices are equal, 
$t_i = t$. For typical metals, $t$ is of the order of 1eV.
Furthermore, the bonding in the gold atomic wire is almost two times stronger than
in the bulk\cite{spain1}.
To simulate the homogeneous monovalent atomic wires (Au, Na, Ag, Cu),
we take some intermediate value of $\Gamma_{L,R}/t = 0.7$ in Fig.~\ref{nntprob}. 
It is a reasonable approximation to assume that the bonding between the 
atom in the atomic wire and the atom in the leads is intermediate 
compared to the bonding in the chain or in the bulk. 
Clearly, $T(\epsilon)$ is peaked at the Fermi energy in the odd-numbered 
atomic wires and is suppressed at $\epsilon= 0$ in the even-numbered 
atomic wires. The number of peaks in $T(\epsilon)$ agrees with the number of 
atoms in the atomic wire.

 Sharp asperity at the STP tip seems to be necessary for the 
conductance quantization according to recent numerical works\cite{spain2}. 
In our approach, the blunt STM tip can be simulated by increasing
the coupling strength between the electrodes and the atomic wire. 
Note that the left (right)-most atom in the chain is coupled to more
atoms in the leads for blunt tips than for sharp tips. 
The results for $\Gamma_{L,R}/t = 1.2$ are displayed in Fig.~\ref{nnbtprob}.
Though the spectral shape of $T_N(\epsilon)$ is broadened compared to 
Fig.~\ref{nntprob}, the even-odd parity effects persist in this case, too. 
Note that the dip of $T_{2N}(\epsilon)$ at the Fermi energy for $t>\Gamma_{L,R}$
is transformed into a peak when $t\leq \Gamma_{L,R}$.

 In the heterogeneous carbon atomic wires\cite{lang}, charge transfer
occurs from the metallic electrode into the carbon chain,
and a Schottky-like barrier is formed between the electrodes and 
the carbon chain.  
Due to the Schottky-like barrier, the coupling strength between the electrodes and 
the chain is weaker than in the homogeneous systems.  
The authors suggest that the reduced transmission probability, $T(0) < 1$,
is a result of the higher barrier between the chain and the leads. 
Higher barrier or weaker coupling leads to sharper line shape in the 
transmission probability. 
We would like to propose a different interpretation. 
As long as the couplings to two electrodes are symmetrical and all 
the hopping matrices are the same, the broken particle-hole symmetry in the 
energy levels of atoms in the chain can be the reason for the reduced $T(0) < 1$.
The doped excess charge is not distributed uniformly over C atoms in the 
chain\cite{lang}. 
The assumption of $E_i = 0$ for all atomic sites seems to be violated.
According to the Friedel sum rule\cite{friedel}, excess (depleted) charges will
shift downward (upward) the energy levels $E_i$ of atoms in the chain.
Broken particle-hole symmetry in the energy structure
leads to reduced transmission probability at the Fermi energy in 
the tight-binding Hamiltonian approach.

 We can also study the effects of disorder in the hopping matrices 
in the atomic wire on the transmission probability.
While retracting the STM tip from the sample, atomic rearrangements 
occur according to molecular dynamics simulations\cite{spain1}. 
Some randomness in the bonding between neighboring atoms might be
present. To simulate this situation, we introduce the disorder
in the magnitude of $t$'s such that $t_i = t(1 + \delta_i)$ with some
randomness $\delta_i$. 
Typical results of $T(\epsilon)$ with disordered
$t_i$'s with $|\delta_i| \leq 0.2$ are displayed in Fig.~\ref{nnrtprob}. 
Even in the presence of disorder in the hopping matrices, the even-odd parity 
behavior may persist as far as the disorder is not strong
and the coupling to the leads is not close to the values of hopping matrices.  

 We predict that the shot noise also displays the even-odd behavior depending 
on the number of atoms in the atomic wire.
First note that $\mbox{Tr} {\bf T}^2 = [T(\epsilon)]^2$ due to 
the simple structure of the linewidth matrices.   
Hence the shot noise is given by the familiar binomial expression\cite{shotnoise} 
and at zero temperature by 
\begin{eqnarray}
S_0 &=& \frac{4e^2}{h} \int_{\mu_R}^{\mu_L} d\epsilon ~ 
 T(\epsilon) [ 1 - T(\epsilon) ]. 
\end{eqnarray}
When the source-drain bias voltage is small such that the transmission probability 
does not change much over the energy $eV=\mu_L - \mu_R$ near the Fermi energy,
the shot noise can be approximated as
$S_0 = (4e^2/h) T(0) [ 1 - T(0)] eV$.    
In the homogeneous sodium atomic wire\cite{kaist}, the transmission probability 
was found to be very close to unity when the number of atoms is odd. 
On the other hand, the transmission probability is smaller than 
unity when the number of atoms is even. We conclude that the shot 
noise at low source-drain bias voltages is larger in the even-numbered atomic wires 
than in the odd-numbered atomic wires.

\section{\label{NS_sec} Normal-superconducting junctions connected by multi-sites}
 In this section, we study the I-V curves in the normal-superconducting(N-S)
junctions connected by multi-sites in between. 
The left lead is a normal metal and the right lead is assumed to be 
the simple BCS $s$-wave superconductors with the constant energy gap.
In our simple approach, the energy gap in the superconducting lead 
is assumed to be constant and not degraded near the junction 
by the atomic wire. Local density of states (DOS) at the atomic sites is 
modified by the superconducting energy gap of the right lead
and its structure depends on the parity in the number of atoms
in the atomic wire. In addition, the DOS structure takes different shapes 
depending on the atom's position in the chain. 
As in the N-N junctions connected by an atomic wire, the Andreev conductance
shows the even-odd parity behavior.

 To describe the superconducting state, it is convenient to 
use the Nambu spinor notation. The two leads are described by the Hamiltonian, 
\begin{eqnarray}
H_{cb}  
 &=& \sum_{p=L,R} \sum_{k} \Psi_{pk}^{\dag} E_{pk} \Psi_{pk}, \\
E_{Lk} &=& \begin{pmatrix} \epsilon_{Lk} & 0 \cr 0 & - \epsilon_{Lk} \end{pmatrix}, ~~
 E_{Rk} = \begin{pmatrix} \epsilon_{Rk} & \Delta_R \cr \Delta_R^{*} & - \epsilon_{Rk}
          \end{pmatrix}, \\
\Psi_{pk} 
 &=& \begin{pmatrix} c_{pk\uparrow} \cr c_{p-k\downarrow}^{\dag} 
     \end{pmatrix}. 
\end{eqnarray}
Here $\epsilon_{pk}$ is the energy dispersion in the left ($p=L$) and 
right ($p=R$) leads and $\Delta_R$ is the energy gap in the right superconducting 
lead. Since the phase of the energy gap $\Delta_R$ in the N-S junction
does not play any role, $\Delta_R$ is taken to be real. 
The Hamiltonian of the multi-sites is also written 
using the Nambu spinor notation as
\begin{eqnarray}
H_d &=& \sum_{i} \Psi_{di}^{\dag} 
    \begin{pmatrix}E_i & 0 \cr 0 & -E_i  \end{pmatrix} \Psi_{di} \nonumber\\
 && + \sum_{i\neq j} \Psi_{di}^{\dag} 
    \begin{pmatrix} W_{ij} & 0 \cr 0 & - W_{ij}^{*} \end{pmatrix} \Psi_{dj},
\end{eqnarray}
where $\Psi_{di}^{\dag} = \begin{pmatrix} d_{i\uparrow}^{\dag} & d_{i\downarrow}
\end{pmatrix}$. $E_i$ is the energy level of the $i$-th site and $W_{ij}$ is
the hopping matrix between two sites $i$ and $j$. The coupling between two leads and the
multi-sites is described by the tunneling Hamiltonian, 
\begin{eqnarray}
H_1 &=& \sum_{pki} \left[ \Psi_{pk}^{\dag} 
     \begin{pmatrix} V_{pi} & 0 \cr 0 & - V_{ip} 
     \end{pmatrix} \Psi_{di}  + H.c.  \right].                      
\end{eqnarray}
Using this model Hamiltonian, we study the Andreev reflection\cite{andreev}
 in the N-S junction.

\subsection{Multi site: General formalism}
 The algebra is highly simplified when the Hamiltonian of the multi-site system 
is written in terms of the matrix. Introducing new notations
\begin{eqnarray}
V_p^{\dag} 
 &\equiv& \begin{pmatrix} V_{p1} & 0 & V_{p2} & 0 & \cdots & V_{pN} & 0 \cr
             0 & V_{1p} & 0 & V_{2p} & \cdots & 0 & V_{Np} \end{pmatrix},    \\
{\bf H}_{d,ii} 
 &=& \begin{pmatrix} E_i & 0 \cr 0 & - E_i \end{pmatrix}, ~~
 {\bf H}_{d,ij} ~=~ \begin{pmatrix} W_{ij} & 0 \cr 0 & - W_{ji} \end{pmatrix}, \\
\Psi_d^{\dag}
 &=& \begin{pmatrix} \Psi_{d1}^{\dag} & \Psi_{d2}^{\dag} & \cdots
             \Psi_{dN}^{\dag} \end{pmatrix},        
\end{eqnarray}
the model Hamiltonian can be written as
\begin{eqnarray}
H_d &=& \Psi_d^{\dag} H_d \Psi_d, \\
H_1 &=& \sum_{pk} \left[ \Psi_{pk}^{\dag} \tau_3 V_p^{\dag} \Psi_d 
         + \Psi_d^{\dag} V_p \tau_3 \Psi_{pk}  \right].
\end{eqnarray}
Here $\tau_3$ is the third component of the Pauli matrices. 
The current operator can be also written in a simple compact form 
using the new notation, 
\begin{eqnarray}
\hat{I}_p 
 &=& \frac{e} {i\hbar} \sum_{k} \left[ \Psi_{pk}^{\dag} V_p^{\dag} \Psi_d 
      - H.c. \right].  
\end{eqnarray}
The thermally averaged current is then 
\begin{eqnarray}
I_p &=& 2e \sum_{k} \mbox{Im} \mbox{Tr} V_p^{\dag} G_{dp}^{<} (t, kt). 
\end{eqnarray}
The mixed Green's function is defined and determined by the Dyson equation, 
\begin{eqnarray}
G_{dp}(t,kt') 
 &=& \frac{1}{i\hbar} \langle T_c \Psi_d(t) \Psi_{pk}^{\dag} (t') \rangle
    \nonumber \\
 &=& \int_C dt_1 ~ D(t,t_1) V_p \tau_3 G_{cp} (k; t_1, t'), \\
i\hbar D(t,t') 
 &=& \langle T_c \Psi_d(t) \Psi_d^{\dag} (t') \rangle. 
\end{eqnarray}
Inserting all the relevant Green's functions of two leads, the current 
can be expressed as 
\begin{widetext}
\begin{eqnarray}
\label{nscurrent}
I_L &=& \frac{4e}{h} \int d\epsilon \mbox{Tr} \left[  
    D^r \cdot \pi N_L V_L \hat{f}_L V_L^{\dag} \cdot D^a 
       \cdot \pi N_L V_L \tau_3 V_L^{\dag} 
   -  D^r \cdot \pi N_L V_L \tau_3 \hat{f}_L V_L^{\dag} \cdot D^a 
       \cdot \pi N_L V_L V_L^{\dag}  \right]   \nonumber\\
 && + \frac{4e}{h} \int d\epsilon \mbox{Tr} 
    \pi N_R g_2(\epsilon) V_R\tau_3 \hat{\Omega}_R \tau_3 V_R^{\dag} 
    \left[ D^r \cdot \pi N_L V_L \tau_3 \hat{f}_L V_L^{\dag} \cdot D^{a} 
       - D^{a} \cdot \pi N_L V_L \tau_3 f(\epsilon) V_L^{\dag} D^{r} 
    \right].
\end{eqnarray}
\end{widetext}
This equation of current is our starting point for the study of the even-odd
parity behavior in the Andreev conductance in the metal-atomic wire-superconducting
junctions. $D^{r,a}$ is the retarded (advanced) part of the Green's function 
of the multi-site system. Note that $D^{a} = [D^{r}]^{\dag}$. 
Other notations are listed below.  
\begin{eqnarray}
\hat{f}_L (\epsilon)
 &=& \begin{pmatrix} f(\epsilon-\mu_L) & 0 \cr 0 & f(\epsilon+\mu_L) 
                 \end{pmatrix}, \\
\hat{\Omega}_R
 &=& \begin{pmatrix} \epsilon & \Delta_R \cr \Delta_R^{*} & \epsilon  
     \end{pmatrix}, \\ 
g(\epsilon) 
 &=& \int \frac{d\epsilon_k} {\pi} \frac{1} {(\epsilon+i\delta)^2 - \epsilon_k^2 
      - |\Delta_R|^2} \nonumber\\
 &=& - \left[ \frac{\theta(\Delta_R - |\epsilon|)} 
                   {\sqrt{\Delta_R^2- \epsilon^2} } 
         + i \mbox{sgn}(\epsilon) \frac{\theta(|\epsilon|-\Delta_R) }
                                     {\sqrt{\epsilon^2-\Delta_R^2} } 
       \right].
\end{eqnarray}
The second line in $g$ is obtained in a wide conduction band limit. 
We write $g = g_1 + i g_2$. Note that $g$ is real for $|\epsilon| < \Delta_R$ and 
imaginary for $|\epsilon| > \Delta_R$.  
To simplify the algebra, the bias voltages are chosen as 
$\mu_L = eV$ and $\mu_R = 0$. 
The first line in Eq.~(\ref{nscurrent})
is a contribution from the Andreev reflection\cite{andreev}. 
The integrand in the first line remains finite over the entire
energy range, though 
it is appreciable inside the gap. 
The electron incident from the normal metallic lead forms a Cooper 
pair with its partner, tunnels into the superconducting lead 
and leave a hole in the normal metallic lead. 
This remaining hole moves along the time-reversed track of the incident electron.  
Note that $g_2(\epsilon) = 0$ for $|\epsilon| < \Delta_R$.  
That is, the second line  in Eq.~(\ref{nscurrent})
is the contribution from quasiparticles excited over 
the superconducting energy gap. The above expression of current 
can be applied to any configuration of multiple resonant sites 
connected to the two leads.

\subsection{One site}
 To begin we consider the N-S junction connected by the one-site system.
The coupling matrices to the leads are
\begin{eqnarray}
V_L^{\dag} 
 &=& \begin{pmatrix} V_{L1} & 0 \cr 0 & V_{1L} \end{pmatrix}, ~~
V_R^{\dag} 
 ~=~ \begin{pmatrix} V_{R1} & 0 \cr 0 & V_{1R} \end{pmatrix}.
\end{eqnarray}
For the one-site system, the Eq.~(\ref{nscurrent}) becomes simplified to 
\begin{eqnarray}
I_L &=& \frac{4e} {h} \Gamma_L^2 \int d\epsilon \mbox{Tr} \left[ 
   D^r \hat{f}_L D^a \tau_3 - D^r \tau_3 \hat{f}_L D^a \right] \nonumber\\
 && + \frac{4e}{h} \Gamma_L \Gamma_R \int d\epsilon g_2(\epsilon) 
   \mbox{Tr} \tau_3 \hat{\Omega} \tau_3   \nonumber\\
 && \hspace{1.0cm} \times \left[ 
    D^r \tau_3 \hat{f}_L D^a - D^a \tau_3 f(\epsilon) D^r \right].
\end{eqnarray}
Here $\Gamma_p$ ($p=L,R$) is the linewidth parameter defined by the relation,
$\Gamma_p = \pi N_p |V_{p1}|^2$.
The first line is the contribution from the Andreev reflection and 
the second line is the contribution from quasiparticles excited 
over the superconducting energy gap. 
The self-energy of a resonant site is given by the equation,
\begin{eqnarray}
\Sigma^{r} 
 &=& - i \Gamma_L {\bf 1} + \Gamma_R g(\epsilon) \tau_3 \hat{\Omega}_R \tau_3, \\
\Sigma^{a} 
 &=& i \Gamma_L {\bf 1} + \Gamma_R g^*(\epsilon) \tau_3 \hat{\Omega}_R \tau_3.
\end{eqnarray}
Here ${\bf 1}$ is a $2\times 2$ unit matrix. 
Inserting the Green's functions of a resonant site, the current can
be written in terms of the transmission probability $T(\epsilon)$,
\begin{eqnarray}
I_L &=& \frac{2e} {h} \int d\epsilon T(\epsilon) 
  \left[ f(\epsilon) - f(\epsilon-\mu_L) \right].
\end{eqnarray} 
$T(\epsilon)$ is defined as a sum of two,
$T(\epsilon) = T_{A} (\epsilon) + T_{qp} (\epsilon)$. 
The Andreev reflection gives 
\begin{eqnarray}
T_A (\epsilon) 
 &=& 8 \Gamma_L^2 |D_{12}^{r} (\epsilon) |^2 \nonumber\\
 &=& \frac{8\Gamma_L^2 \Gamma_R^2} {|Z_1(\epsilon)|^2} 
    \frac{\Delta_R^2} {|\epsilon^2 - \Delta_R^2|} 
 ~=~ \frac{8\Gamma_L^2 \Gamma_R^2 \Delta_R^2} {{\cal D} (\epsilon)}.
\end{eqnarray}
Here $Z_1$ is the determinant of $[D^{r}(\epsilon)]^{-1}$ and
the denominator ${\cal D}$ is given by the expressions, 
\begin{eqnarray}
{\cal D}
 &=& \left[ (\epsilon^2 - E_1^2 - \Gamma_L^2 - \Gamma_R^2) 
        \sqrt{\Delta_R^2 - \epsilon^2} + 2 \Gamma_R \epsilon^2 \right]^2  \nonumber\\
 && + 4\Gamma_L^2 \epsilon^2 \left[ 
         \Gamma_R + \sqrt{\Delta_R^2 - \epsilon^2} \right]^2, 
\end{eqnarray}
for $|\epsilon| < \Delta_R$ and 
\begin{eqnarray}
{\cal D}
 &=& \left[ (\epsilon^2 - E_1^2 - \Gamma_L^2 - \Gamma_R^2) 
       \sqrt{\epsilon^2 - \Delta_R^2} - 2\Gamma_L \Gamma_R |\epsilon| \right]^2
      \nonumber\\
 && + 4\epsilon^2 \left[ \Gamma_L \sqrt{\epsilon^2 - \Delta_R^2} 
           + \Gamma_R |\epsilon| \right]^2, 
\end{eqnarray}
for $|\epsilon| > \Delta_R$. 
Though the detailed forms of $T_A(\epsilon)$ are different depending on 
the energy range, the Andreev contribution is continuous 
at $|\epsilon| = \Delta_R$.  
\begin{eqnarray}
T_A(0) 
 &=& \frac{8\Gamma_L^2 \Gamma_R^2} {(E_1^2 + \Gamma_L^2 + \Gamma_R^2)^2}, \\
T_A(\pm\Delta_R) 
 &=& \frac{2\Gamma_L^2} {\Gamma_L^2 + \Delta_R^2}. 
\end{eqnarray}
Note that $T_A$ at $\epsilon = \pm \Delta_R$ is independent of the energy
level position $E_1$ and is always less than a value of $2$.  
The quasiparticle contribution is given in an explicit form as
\begin{widetext}
\begin{eqnarray}
T_{qp} (\epsilon) 
 &=& 4\Gamma_L \Gamma_R |g_2(\epsilon)| \left[ 
   |\epsilon| \left\{ |D_{11}^{r}(\epsilon)|^2 + |D_{12}^{r}(\epsilon)|^2 \right\} 
  - \Delta_R \mbox{sgn}(\epsilon) \mbox{Re} \left\{ 
     D_{11}^{r} (\epsilon) D_{12}^{a}(\epsilon) 
     - D_{22}^{r} (-\epsilon) D_{12}^{a} (-\epsilon) \right\} 
  \right] \nonumber\\
 &=& \frac{ 4\Gamma_L \Gamma_R  \sqrt{\epsilon^2 - \Delta_R^2} 
    \left[  |\epsilon| 
     \left\{ (\epsilon+E_1)^2 + \Gamma_L^2 + \Gamma_R^2 \right\}
     + 2\Gamma_L \Gamma_R \sqrt{\epsilon^2 - \Delta_R^2} \right] } 
   { \left[ (\epsilon^2 - E_1^2 - \Gamma_L^2 - \Gamma_R^2) 
            \sqrt{\epsilon^2 - \Delta_R^2} - 2\Gamma_L \Gamma_R |\epsilon| \right]^2
     + 4\epsilon^2 \left[ \Gamma_L \sqrt{\epsilon^2 - \Delta_R^2} 
           + \Gamma_R |\epsilon| \right]^2   }.  
\end{eqnarray}
\end{widetext}
As expected, the quasiparticle contribution vanishes at $|\epsilon| = \Delta_R$.  
When the right lead becomes normal or $\Delta_R = 0$, the Andreev contribution
vanishes and the quasiparticle contribution is simplified to the well-known 
Lorentzian form, 
$T_{qp} = 4\Gamma_L \Gamma_R / [(\epsilon - E_1)^2 + (\Gamma_L + \Gamma_R)^2]$.

\subsection{Atomic wire: serial multi-sites}
When multi-sites are arranged in series like atomic wires, 
the current can be simplified considerably. 
Since the left (right)-most site is coupled to the left (right) lead, respectively,
the coupling matrices to the leads are given by the expressions,
\begin{eqnarray}
V_L^{\dag} 
 &=& \begin{pmatrix} \begin{matrix} V_{L1} & 0 \cr 0 & V_{1L} \end{matrix} &
           {\bf 0} & \cdots & {\bf 0} \end{pmatrix}, \\
V_R^{\dag} 
 &=& \begin{pmatrix} {\bf 0} & \cdots {\bf 0} & 
      \begin{matrix} V_{RN} & 0 \cr 0 & V_{NR} \end{matrix} \end{pmatrix}.
\end{eqnarray}
Here ${\bf 0}$ is the $2\times 2$ zero matrix. 
Various matrices appearing in the expression of current are also simplified.
For example,  
\begin{eqnarray}
~[\pi N_L V_L V_L^{\dag}]_{ij} 
 &=& \Gamma_L {\bf 1} \delta_{i,1}\delta_{j,1}.
\end{eqnarray}
Here ${\bf 1}$ is the $2\times 2$ unit matrix in the Nambu spinor space. 
The Andreev contribution
to current can be written in components as
\begin{eqnarray}
I_A &=& \frac{4e}{h} \int d\epsilon \mbox{Tr} \left[ 
   {\bf D}_{11}^{r} \Gamma_L \hat{f}_L {\bf D}_{11}^{a} \Gamma_L \tau_3 
   - {\bf D}_{11}^{r} \Gamma_L \hat{f}_L \tau_3 {\bf D}_{11}^{a} \Gamma_L \right]
   \nonumber\\
 &=& \frac{8e}{h} \cdot \Gamma_L^2 \int d\epsilon 
  |D_{11,12}^{r} (\epsilon) |^2 \left[ f(\epsilon+\mu_L) - f(\epsilon-\mu_L) \right].
\end{eqnarray}
The indices $i$ and $j$ in $D_{ij, \alpha\beta}$ denote the sites of resonant levels
while the indices $\alpha$ and $\beta$ mean the elements of the $2\times 2$ matrix
in the Nambu spinor space. 
\begin{eqnarray}
I_A &=& \frac{2e}{h} \int d\epsilon T_A(\epsilon) 
    \left[ f(\epsilon) - f(\epsilon-\mu_L) \right], \\
T_A(\epsilon)
 &=& 4\Gamma_L^2 \left[ |D_{11,12}^{r} (\epsilon) |^2
     + |D_{11,12}^{r} (-\epsilon) |^2 \right]. 
\end{eqnarray}
The Andreev conductance is given in a simple form, 
\begin{eqnarray}
G_{NS} &=& 2 \cdot \frac{2e^2}{h} \cdot 4\Gamma_L^2 | D_{11,12}^{r} (0) |^2. 
\end{eqnarray}
The quasiparticle contribution to current 
is also given in components by the equation,
\begin{eqnarray}
I_{qp}
 &=& - \frac{4e}{h} \int d\epsilon ~\Gamma_L \overline{\Gamma}_R(\epsilon) 
    \mbox{sgn}(\epsilon) \mbox{Tr} \tau_3 \hat{\Omega}_R \tau_3 \nonumber\\
 && \times
   \left[ {\bf D}_{N1}^r \hat{f}_L \tau_3 {\bf D}_{1N}^{a} 
        - {\bf D}_{N1}^a f \tau_3 {\bf D}_{1N}^{r} 
    \right].
\end{eqnarray}
Here $\overline{\Gamma}_R(\epsilon) = \Gamma_R |g_2(\epsilon)|$.  
After some algebra, the quasiparticle contribution to the current can be 
written as
\begin{widetext}
\begin{eqnarray}
I_{qp} &=& \frac{2e}{h} \int d\epsilon T_{qp} (\epsilon) 
  \left[ f(\epsilon) - f(\epsilon - \mu_L) \right], \\ 
T_{qp} (\epsilon) 
 &=& 2\Gamma_L \overline{\Gamma}_R(\epsilon) 
    |\epsilon| \left[ |D_{N1,11}^{r} (\epsilon)|^2 
          + |D_{N1,12}^{r} (\epsilon)|^2 
          + |D_{N1,22}^{r} (-\epsilon)|^2 
          + |D_{N1,12}^{r} (-\epsilon)|^2 \right]  \nonumber\\
 && - 4 \Gamma_L \overline{\Gamma}_R(\epsilon)  
     \Delta_R \mbox{sgn}(\epsilon) ~ \mbox{Re} \left[ 
     D_{N1,11}^{r} (\epsilon) D_{1N, 12}^{a} (\epsilon) 
     - D_{N1,22}^{r} (-\epsilon) D_{1N, 21}^{a} (-\epsilon) \right].
\end{eqnarray}
\end{widetext}
The total transmission probability is a sum of two contributions: 
$T = T_A + T_{qp}$. Though we are using the word ``probability", 
it is a misnomer
because $T$ can become larger than a unity due to the Andreev reflection 
below the superconducting energy gap.

 In a serial multi-site configuration, the retarded self-energy is given by the 
expression, 
\begin{eqnarray}
\Sigma^{r}_{ij} 
 &=& -i \Gamma_L {\bf 1} \delta_{i,1} \delta_{j,1}
   + \Gamma_R g(\epsilon) \tau_3 \hat{\Omega}_R \tau_3 ~ \delta_{i,N} \delta_{j,N},
\end{eqnarray}
and the retarded Green's function of multi-site system is expressed by the 
matrix inversion, $D^r = [ \epsilon - H_d - \Sigma^{r}]^{-1}$.
We now calculate the Andreev conductance of multi-site atomic wire one by one.
For one resonant site, the Green's function is given by the equation, 
\begin{eqnarray}
D^{r} (\epsilon)
 &=& \begin{pmatrix} \epsilon Z_R - E_1 + i\Gamma_L
              & \Gamma_R \Delta_R g(\epsilon) \cr
     \Gamma_R \Delta_R g(\epsilon) & 
        \epsilon Z_R + E_1 + i\Gamma_L 
   \end{pmatrix}^{-1}, \\
G_{NS} &=& 2\cdot \frac{2e^2}{h} \cdot \frac{4\Gamma_L^2 \Gamma_R^2} 
    {[E_1^2 + \Gamma_L^2 + \Gamma_R^2]^2}.  
\end{eqnarray}
Here $Z_R = 1 - \Gamma_R g(\epsilon)$ and the one-site system 
is detailed in the previous section. 
For two-site atomic wire, the Green's function relevant to the Andreev 
reflection is 
\begin{eqnarray}
D_{11,12}^{r}(\epsilon)
 &=& \frac{|t_1|^2 \Delta_R \Gamma_R g(\epsilon)} {Z_2(\epsilon)}, 
\end{eqnarray}
where
\begin{eqnarray}
Z_2(\epsilon) 
 &=& [ (\epsilon+i\Gamma_L)^2 - E_1^2] 
   [\epsilon^2 - E_2^2 - \Gamma_R^2 - 2\epsilon^2 \Gamma_R g ] \nonumber\\
 && -2 |t_1|^2 \left[ (\epsilon + i\Gamma_L)
                            (\epsilon - \epsilon \Gamma_R g) 
             + E_1 E_2  \right]  \nonumber\\
 && + |t_1|^4.
\end{eqnarray}
The Andreev conductance is 
\begin{eqnarray}
G_{NS} &=& 2 \cdot \frac{2e^2}{h} \cdot 
   \frac{4\Gamma_L^2\Gamma_R^2 |t_1|^4} {|Z_1(0)|^2}, \\
Z_2(0) 
 &=& (E_1^2 + \Gamma_L^2) ( E_2^2 + \Gamma_R^2) + |t_1|^4 \nonumber\\
 && - 2|t_1|^2 E_1 E_2.
\end{eqnarray}
At $\epsilon = \pm \Delta_R$, $T_A$ is independent of $E_2$ 
and is given by the expression, 
\begin{eqnarray}
T_A
 &=& \frac{ 2 \Gamma_L^2 t_1^4 } 
   {\Delta_R^2 (E_1^2 + \Gamma_L^2 + t_1^2 - \Delta_R^2)^2
     + \Gamma_L^2 (2\Delta_R^2 - t_1^2)^2 }. 
\end{eqnarray}
$T_A$ at $\epsilon = \pm\Delta_R$ can reach the maximum value of $2$ 
when $E_1=0$ and $\Gamma_L = \sqrt{t^2 - \Delta_R^2}$. 
For three-site atomic wire, the Green's function at the Fermi energy is 
\begin{eqnarray}
D^{r}_{11,12} (0) 
 &=& \frac{ |t_1|^2 |t_2|^2 \Gamma_R} {Z_3(0)},
\end{eqnarray}
where 
\begin{eqnarray}
Z_3(0) 
 &=& \mbox{det} ( - {\bf H}_d + i{\bf \Gamma} )  \nonumber\\
 &=& - (E_1^2 + \Gamma_L^2) E_2^2 (E_3^2 + \Gamma_R^2) 
   - |t_2|^4 (E_1^2 + \Gamma_L^2) \nonumber\\
 &&  - |t_1|^4 (E_3^2 + \Gamma_R^2) 
   + 2E_2 E_3 |t_2|^2 (E_1^2 + \Gamma_L^2) \nonumber\\
 && + |t_1|^2 \left\{ E_1 E_2 (2E_3^2 + \Gamma_L^2) - E_1 E_3 |t_2|^2 \right\}
       \nonumber\\
 && - i |t_1|^2 \Gamma_L (E_2 \Gamma_L^2 - E_3 |t_2|^2).
\end{eqnarray}
When $E_i = 0$ for all $i$'s, the Andreev conductance is given in a simple 
form,   
\begin{eqnarray}
D_{11, 12}^{r} (0) 
 &=& - \frac{ |t_1|^2 |t_2|^2 \Gamma_R } { \Gamma_L^2 |t_2|^4 + \Gamma_R^2 |t_1|^4}, \\
G_{NS} &=& 2 \cdot \frac{2e^2}{h} \cdot 
  \frac{ 4\Gamma_L^2 |t_2|^4 \cdot \Gamma_R^2 |t_1|^4 } 
    { \left[ \Gamma_L^2 |t_2|^4 + \Gamma_R^2 |t_1|^4  \right]^2 }.
\end{eqnarray}
We can now deduce the general expression of the Andreev conductance for $N$-site
atomic wire. Depending on the parity of the number of atoms in the atomic wire, 
the Andreev conductance takes different forms,
\begin{eqnarray}
G_{NS} &=& \frac{2e^2}{h} \cdot T_N(0), \\
T_1 (0) &=& \frac{8 \cdot \Gamma_L^2 \cdot \Gamma_R^2} 
           { \left[ \Gamma_L^2 + \Gamma_R^2 \right]^2 }, \\
T_{2N+1}(0)
 &=& \frac{8 \cdot \Gamma_L^2 |t_2 t_4 \cdots t_{2N}|^4 \cdot 
              \Gamma_R^2 |t_1 t_3 \cdots t_{2N-1}|^4 }
       { \left[ \Gamma_L^2 |t_2 t_4 \cdots t_{2N}|^4  
           + \Gamma_R^2 |t_1 t_3 \cdots t_{2N-1}|^4 \right]^2 }, \\
T_2(0) 
 &=& \frac{8 \cdot \Gamma_L^2\Gamma_R^2 \cdot |t_1|^4}
     { \left[ \Gamma_L^2 \Gamma_R^2 + |t_1|^4 \right]^2}, \\
T_{2N}(0) 
 &=& \frac{8 \cdot \Gamma_L^2 |t_2 t_4 \cdots t_{2N-2}|^4 \Gamma_R^2  \cdot 
           |t_1 t_3 \cdots t_{2N-1}|^4 } 
    { \left[ \Gamma_L^2 |t_2 t_4 \cdots t_{2N-2}|^4 \Gamma_R^2
         +  |t_1 t_3 \cdots t_{2N-1}|^4 \right]^2 }.
\end{eqnarray}
The even-odd behavior is also evident in the Andreev conductance
as in the NAN junction connected by the atomic wire. 
Since the Andreev reflection is the two-particle process, 
all the parameters in the transmission probability at $E_F$ 
in the NAS junction are squared compared to the NAN junction.  
In the mirror symmetric atomic wire, $T_{2N+1}(0)$ does not 
depend on the detailed values of the hopping matrices in the atomic chain
but is determined by the couplings to the leads.
\begin{eqnarray}
T_{2N+1}(0)
 &=& \frac{8 \cdot \Gamma_L^2  \cdot \Gamma_R^2 }
       { \left[ \Gamma_L^2 + \Gamma_R^2 \right]^2 }.
\end{eqnarray} 
$T_{2N}(0)$ depends on the hopping matrices in the atomic wire. 
When all the hopping integrals are equal (a special case of the mirror 
symmetric chain), the transmission probability at the Fermi energy is
simplified as
\begin{eqnarray}
T_{2N+1}(0)
 &=& \frac{8 \cdot \Gamma_L^2  \cdot \Gamma_R^2 }
       { \left[ \Gamma_L^2 + \Gamma_R^2 \right]^2 }, \\
T_{2N}(0)
 &=& \frac{8 \cdot \Gamma_L^2 \Gamma_R^2 \cdot t^4}
       { \left[ \Gamma_L^2 \Gamma_R^2 + t^4\right]^2 }.
\end{eqnarray} 
The maximum transmission is achieved when $\Gamma_L = \Gamma_R$ in the 
odd-numbered atomic wire, and when $t^2 = \Gamma_L \Gamma_R$ in the 
even-numbered atomic wire.

 To illustrate the even-odd parity dependence of the transmission probability 
$T_N(\epsilon)=T_A(\epsilon) + T_{qp}(\epsilon)$, we plot $T_N(\epsilon)$ for 
differing $N$'s (see Figs.~\ref{nstprob} and \ref{nsbtprob}). 
The model parameters are chosen as follows:
$E_i = 0$ for all the sites and $t_i = t$ for all $i$'s. 
The superconducting energy gap $\Delta_R = t/10$ is chosen to be large 
to show clearly the Andreev structure in $T_N(\epsilon)$. 
The results for the case $\Gamma_{L,R} = 0.3\times t$
are displayed in Fig.~\ref{nstprob}. 
Note that the Andreev contribution of $T_A(\epsilon)$ is appreciable 
only inside the superconducting energy gap. 
For odd-numbered atomic wires, the transmission probability displays 
the Andreev peak at the Fermi energy and the side peaks above the superconducting
energy gap which are the quasiparticle contributions. 
For even-numbered atomic wires, the transmission probability 
is suppressed at the Fermi energy and is peaked at the superconducting energy gap.
Other peaks outside the superconducting energy gap are contributions from 
the quasiparticles.

 When the coupling to the leads is larger than the hopping integral in the 
atomic wire, the spectral shape of $T_N(\epsilon)$ is different 
compared to the case of $\Gamma_{L,R} < t$. 
The results of $T_N(\epsilon)$ when $\Gamma_{L,R} = 1.2\times t$ 
are displayed in Fig.~\ref{nsbtprob}. 
Increased coupling to the leads results in a broadened spectral shape 
of the transmission probability.  In the odd-numbered atomic wire, 
the transmission probability is almost flat inside the superconducting gap.
In the even-numbered atomic wire, the Andreev conductance is increased 
as $\Gamma_{L,R}$ is increased($t > \sqrt{\Gamma_L\Gamma_R}$), 
reaches the maximum value of 2
when $t = \sqrt{\Gamma_L\Gamma_R}$
and is decreased with further increasing $\Gamma_{L,R}$($\sqrt{\Gamma_L\Gamma_R} > t$).

 The local DOS at each atomic site in the atomic wire is computed and 
displayed in Figs.~\ref{nsdos4} and \ref{nsdos5}. 
The shape of local DOS is 
 dependent on the parity of the number of atoms in the atomic wire.
The local DOS also depends on the atom's position in the chain.

\section{\label{sum_sec} Summary and conclusion}
 Using the tight-binding Hamiltonian, we studied the dependence of 
the conductance and shot noise on the number of atoms in the atomic wire 
when the atomic wire is connected to the two metallic leads or 
to the metal and superconducting leads.
In metal-atomic wire-metal(NAN) junctions, the even-odd parity dependence 
of the conductance can be understood by the energy-level splitting in 
quantum physics and the particle-hole symmetry of atomic energy levels.
The linear response conductance is larger in the odd-numbered atomic 
wire than in the even-numbered atomic wire. 
The conductance can reach the maximum possible value $2e^2/h$ in the 
odd-numbered atomic wire when all the atomic energy levels lie
at the Fermi energy (resonance condition) and 
the system is mirror symmetric with respect to the center of the wire. 
The shot noise also displays the even-odd parity effects depending 
on the number of atoms in the atomic wire. 
In contrast to the conductance, the shot noise is larger in the even-numbered
atomic wire than in the odd-numbered atomic wire.

 We also studied the Andreev conductance in the metal-atomic wire-superconducting
junctions. As in the NAN junctions, the Andreev conductance displays 
the even-odd parity effects. 
In the odd-numbered atomic wire, the conductance is enhanced and 
close to $4e^2/h$ due to the Andreev reflection. 
The maximum Andreev conductance $4e^2/h$ is possible in the mirror-symmetric 
odd-numbered atomic wire when the resonance condition is satisfied.   
On the other hand, 
the conductance is suppressed below $4e^2/h$
in the even-numbered atomic wire even when the resonance condition 
is satisfied for the mirror-symmetric case.

\acknowledgments
We would like to thank H. W. Lee for his careful reading of this manuscript.
This work was supported in part by the National 
Science Foundation under Grant No. DMR 9357474, 
in part by the BK21 project and
in part by grant No. 1999-2-114-005-5 from the KOSEF.

\appendix

\section{\label{app_noise} Derivation of Eq.~(\ref{shoteqn})}
\begin{figure}[h]
\protect\centerline{\epsfxsize=0.40\textwidth \epsfbox{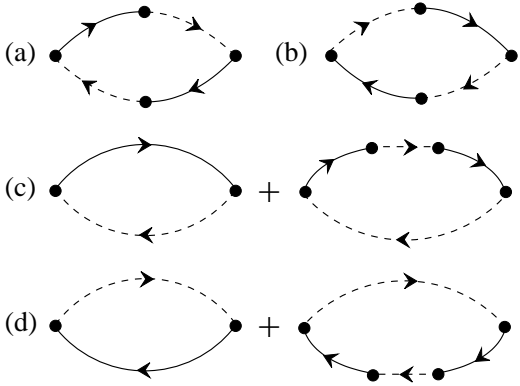}}
\vskip 0.5cm
\caption{Feynman diagrams for the current noise.  
Solid line represents the conduction electron 
Green's function, while the dashed line means the fully dressed Green's function
of the multi-site system.
\label{noise}}
\end{figure}

\begin{figure}[h]
\protect\centerline{\epsfxsize=0.45\textwidth \epsfbox{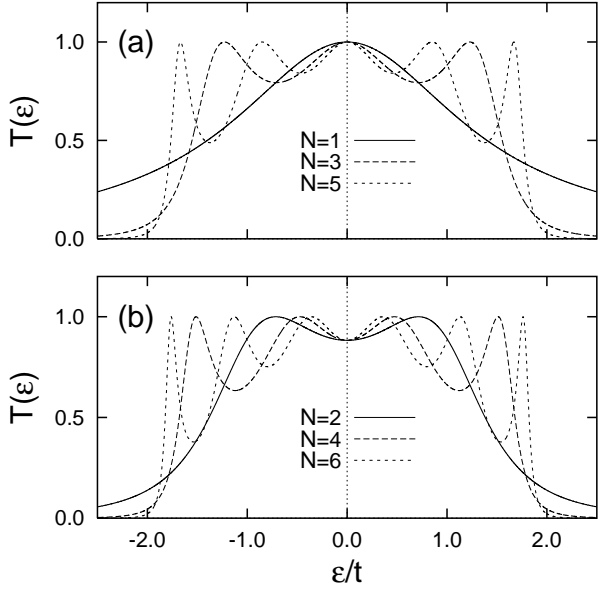}}
\vskip 0.5cm
\caption{Dependence of the transmission probability on the number of 
atoms in atomic wires in the metal-atomic wire-metal junctions. 
Panel (a) displays the transmission probability 
for the odd-numbered atomic wires. $T(\epsilon)$ is peaked at the Fermi energy. 
On the other hand, panel (b) shows the transmission probability 
for the even-numbered atomic wires. $T(\epsilon)$ is suppressed at the 
Fermi energy. We take the model parameters: $E_i = 0$ and $t_i = t$ 
for all sites, $\Gamma_{L,R} = 0.7\times t$. See the text for explanations.  
\label{nntprob}}
\end{figure}

\begin{figure}[h]
\protect\centerline{\epsfxsize=0.45\textwidth \epsfbox{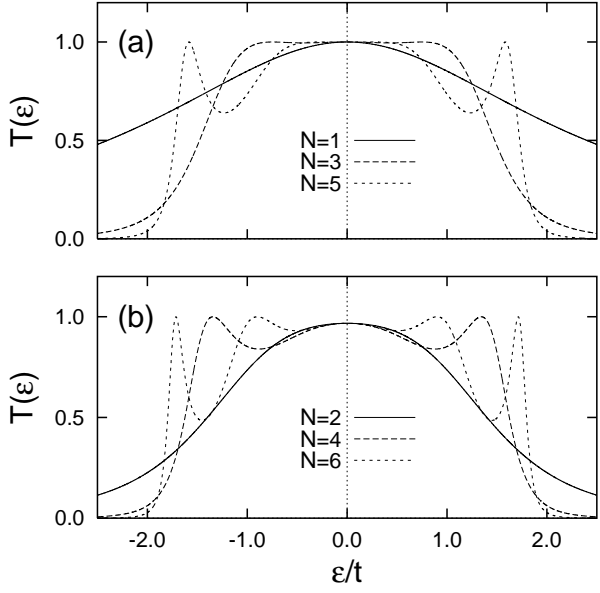}}
\vskip 0.5cm
\caption{The transmission probability for blunt tips. Model parameters 
are the same as in Fig.~\ref{nntprob} except for $\Gamma_{L,R} = 1.2\times t$. 
Increased coupling to the leads broadens the lineshape of the transmission 
probability. $T_{2N}(\epsilon)$ is peaked at $\epsilon=0$ when 
$\Gamma_{L,R} \geq t$.
 \label{nnbtprob}}
\end{figure}

\begin{figure}[h]
\protect\centerline{\epsfxsize=0.45\textwidth \epsfbox{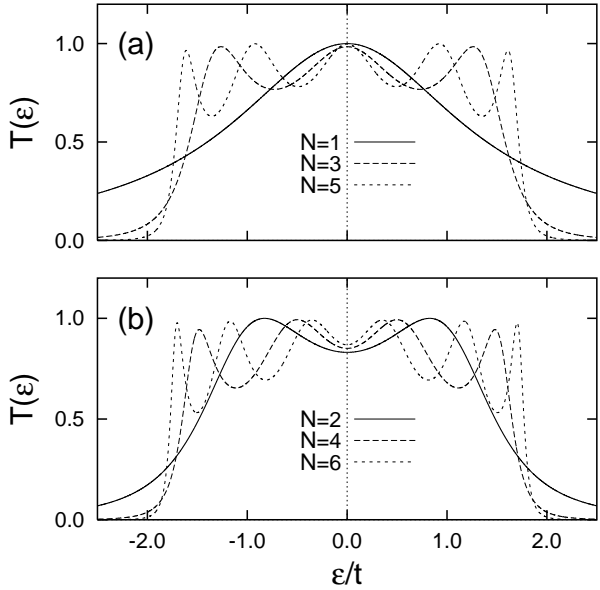}}
\vskip 0.5cm
\caption{The transmission probability for disordered hopping matrices.
 Model parameters 
are the same as in Fig.~\ref{nntprob} except that some disorder in $t_i$'s
is introduced: $t_i = t(1 + \delta_i)$ with $|\delta_i| \leq 0.2$. 
 \label{nnrtprob}}
\end{figure}

\begin{figure}[h]
\protect\centerline{\epsfxsize=0.45\textwidth \epsfbox{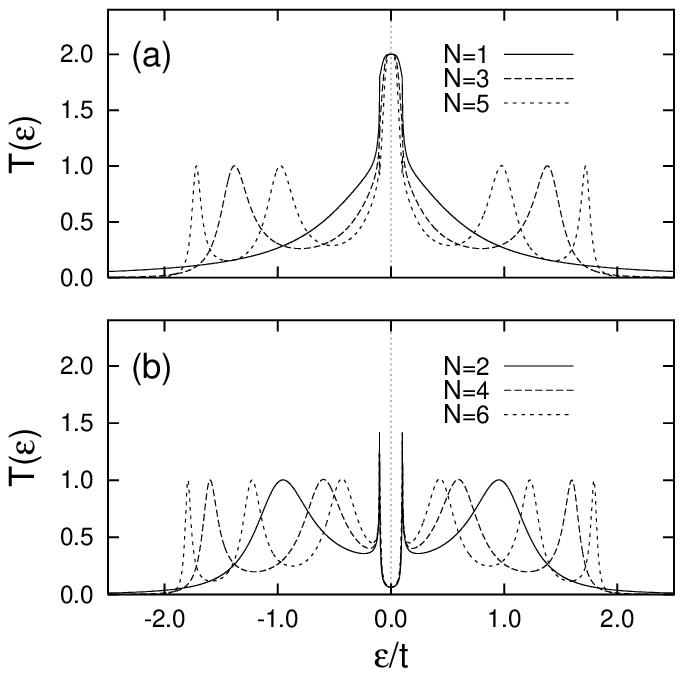}}
\vskip 0.5cm
\caption{Dependence of the transmission probability on the number of 
atoms in an atomic wire in the metal-atomic wire-superconducting junctions.
Panel (a) for the odd-numbered atomic wires and panel (b) for the even-numbered
atomic wires. We take model parameters: $E_i = 0$ and $t_i = t$ for all sites, 
$\Gamma_{L,R} = 0.3\times t$ and the superconducting energy gap
$\Delta_R = 0.1\times t$.
\label{nstprob}}
\end{figure}

\begin{figure}[h]
\protect\centerline{\epsfxsize=0.45\textwidth \epsfbox{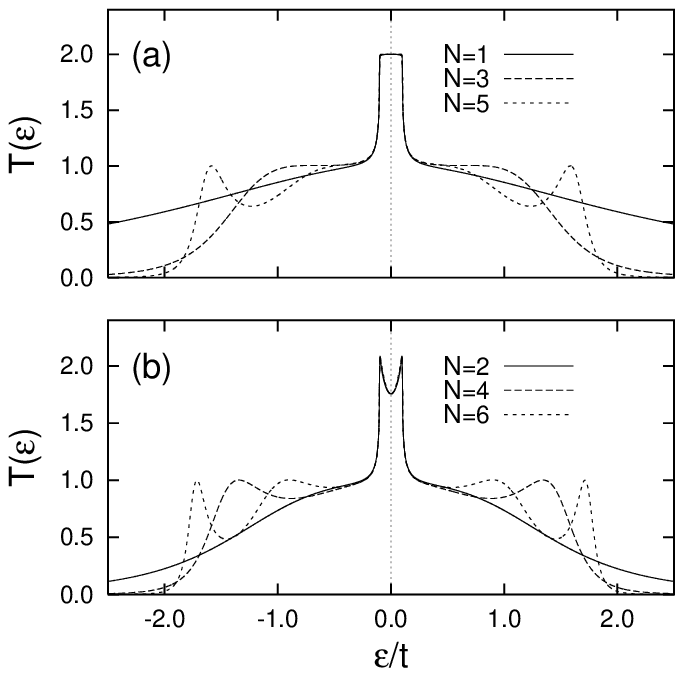}}
\vskip 0.5cm
\caption{Dependence of the transmission probability on the number of 
atoms in an atomic wire in the metal-atomic wire-superconducting junctions.
Panel (a) for the odd-numbered atomic wires and panel (b) for the even-numbered
atomic wires. We take model parameters: $E_i = 0$ and $t_i = t$ for all sites, 
$\Gamma_{L,R} = 1.2\times t$ and the superconducting energy gap
$\Delta_R = 0.1\times t$. 
\label{nsbtprob}}
\end{figure}

\begin{figure}[h]
\protect\centerline{\epsfxsize=0.45\textwidth \epsfbox{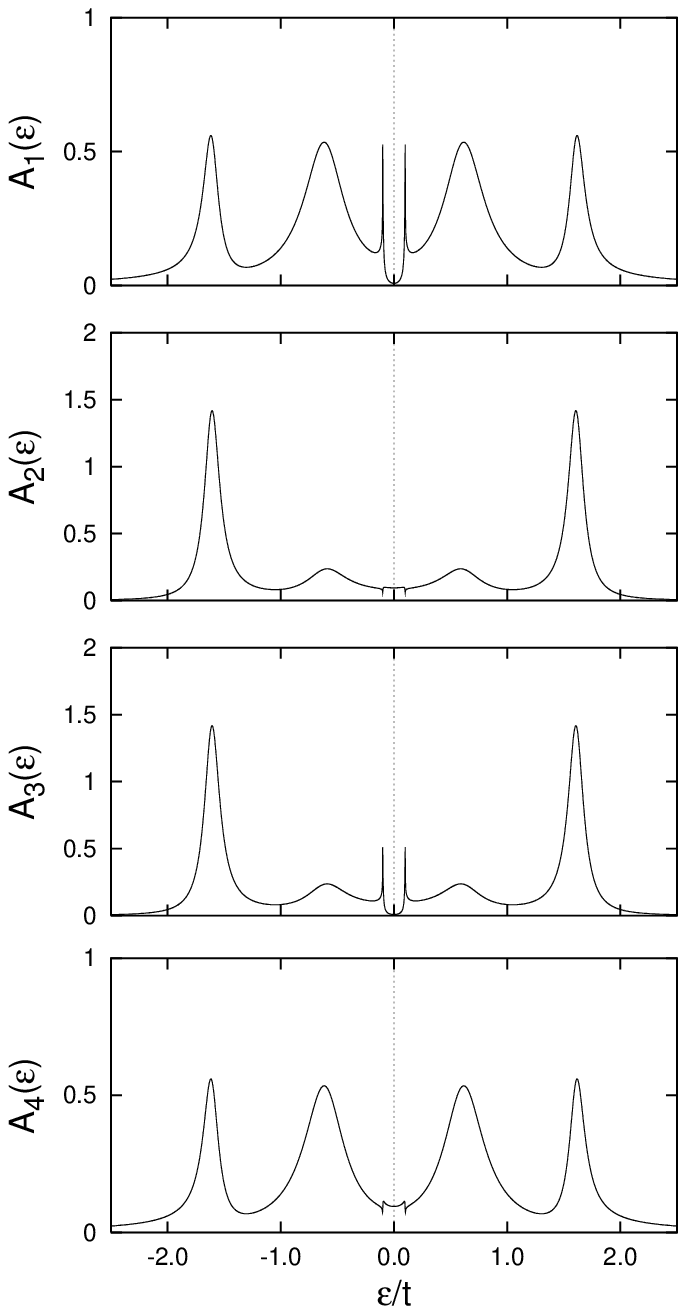}}
\vskip 0.5cm
\caption{Local DOS at each atomic site for 4-site atomic wire. 
$A_i$ means the spectral function of the $i$-th atoms from the left lead (normal 
metals). Model parameters are the same as in Fig.~\ref{nstprob}.
\label{nsdos4}}
\end{figure}

\begin{figure}[h]
\protect\centerline{\epsfxsize=0.45\textwidth \epsfbox{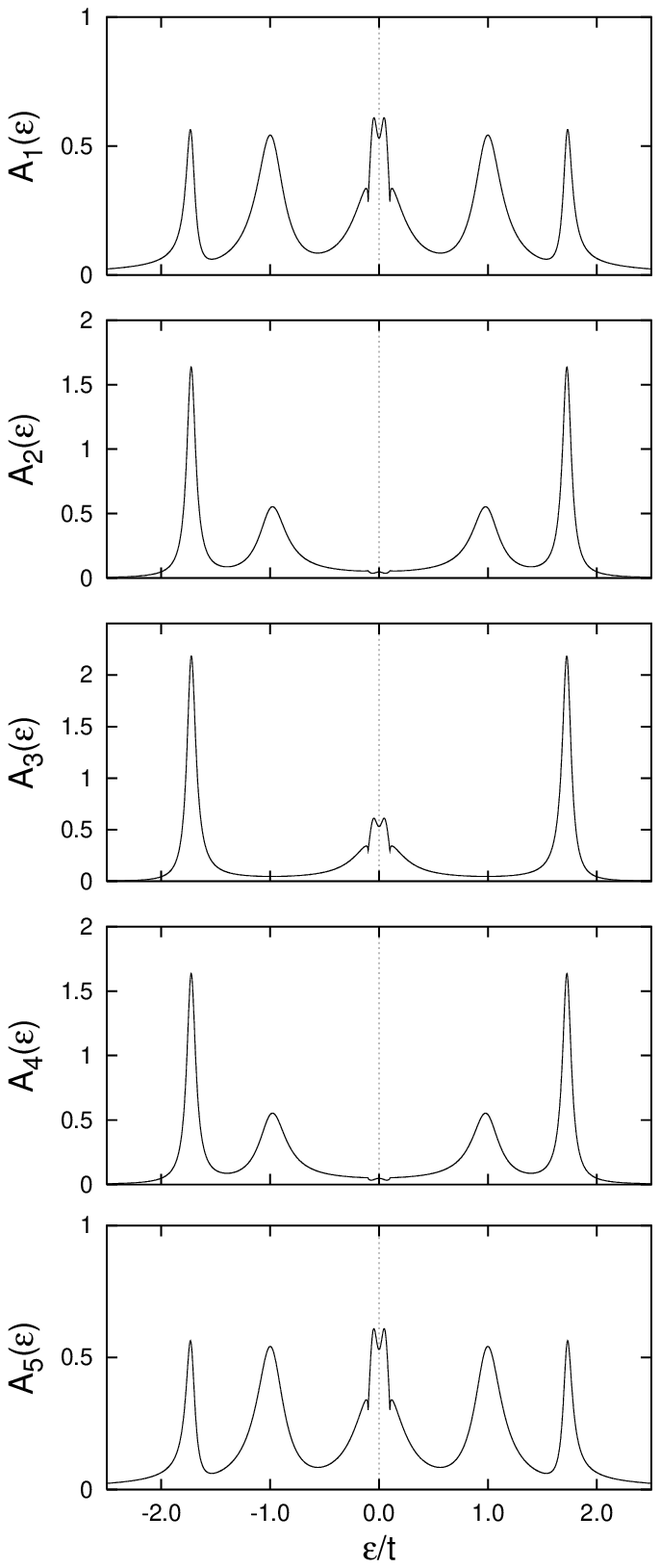}}
\vskip 0.5cm
\caption{Local DOS at each atomic site for 5-site atomic wire.
$A_i$ means the spectral function of the $i$-th atoms from the left lead (normal 
metals). Model parameters are the same as in Fig.~\ref{nstprob}.
\label{nsdos5}}
\end{figure}

 In this Appendix, we find the expression of the current noise
by calculating the Feynman diagrams shown in Fig.~\ref{noise}.  
We calculate each diagram step by step. 
The diagram in Fig.~\ref{noise}(a) gives 
\begin{widetext}
\begin{eqnarray}
i\hbar G_{II} (t,t') 
 &=& - \frac{e^2}{V^2} \sum_{kk'} \int_C dt_1 \int_C dt_2 ~
    \mbox{Tr} \left\{  D(t,t_1) V_L 
     G_{cL} (k'; t_1, t') V_L^{\dag} D(t',t_2) V_L  
     G_{cL} (k; t_2, t) V_L^{\dag}  \right\}.
\end{eqnarray}
After some algebra, the lesser part is given by the equation
\begin{eqnarray}
\hbar G_{II}^{<} (\epsilon) 
 &=& \frac{e^2}{h} \int d\zeta ~ \mbox{Tr} 
   \left\{  \left[ i D^{<} (\epsilon+\zeta) 
          + 2f_L(\epsilon+\zeta) D^{r} (\epsilon+\zeta)  \right]
      \Gamma_L  \left[ i D^{>} (\zeta) 
          + 2\bar{f}_L(\zeta) D^{r} (\zeta)  \right]
      \Gamma_L  \right\}.
\end{eqnarray}
We use the analytic continuation to the real-time axis\cite{langreth} 
to find the lesser current Green's function.  
The diagram in Fig.~\ref{noise}(b) gives 
\begin{eqnarray}
i\hbar G_{II} (t,t') 
 &=& - \frac{e^2}{V^2} \sum_{kk'} \int_C dt_1 \int_C dt_2 ~
    \mbox{Tr} \left\{ 
     V_L G_{cL} (k; t, t_1) V_L^{\dag} D(t_1,t')  
     V_L G_{cL} (k'; t', t_2) V_L^{\dag} D(t_2,t)  \right\}, \\
\hbar G_{II}^{<} (\epsilon) 
 &=& \frac{e^2}{h} \int d\zeta ~ \mbox{Tr} 
   \left\{ \Gamma_L  \left[ - i D^{<} (\epsilon+\zeta) 
          + 2f_L(\epsilon+\zeta) D^{a} (\epsilon+\zeta)  \right]
      \Gamma_L  \left[ -i D^{>} (\zeta) 
          + 2\bar{f}_L(\zeta) D^{a} (\zeta)  \right]
   \right\}.
\end{eqnarray}
Two diagrams in Fig.~\ref{noise}(c) give 
\begin{eqnarray}
i\hbar G_{II} (t,t') 
 &=& \frac{e^2}{V} \sum_{k} 
    \mbox{Tr} \left\{ V_L  G_{cL} (k;t,t') V_L^{\dag}
     D (t', t)  \right\}  \nonumber\\
 && + \frac{e^2}{V^2} \sum_{kk'} \int_C dt_1 \int_C dt_2 ~
    \mbox{Tr} \left\{ V_L G_{cL} (k; t, t_2)
     V_L^{\dag} D(t_2,t_1) V_L 
     G_{cL} (k'; t_1, t') V_L^{\dag}  D(t',t)
     \right\}, \\
\hbar G_{II}^{<} (\epsilon) 
 &=& - \frac{2e^2}{h} \int d\zeta ~ f_L (\epsilon+\zeta) 
     \mbox{Tr} \{ \Gamma_L D^{>} (\zeta)  \} 
    \nonumber\\
 && - \frac{e^2}{h} \int d\zeta ~ \mbox{Tr} 
   \left\{ \Gamma_L
      \left[ D^{<} (\epsilon+\zeta) 
         + 2i f_L (\epsilon+\zeta) \{ D^{a} (\epsilon+\zeta) 
             - D^{r} (\epsilon+\zeta) \}
      \right] \Gamma_L D^{>} (\zeta)
   \right\}.
\end{eqnarray}
Two diagrams in Fig.~\ref{noise}(d) give 
\begin{eqnarray}
i\hbar G_{II} (t,t') 
 &=& \frac{e^2}{V} \sum_{k} 
    \mbox{Tr} \left\{ D(t,t') V_L
     G_{cL} (k; t', t) V_L^{\dag} \right\}  \nonumber\\
 && + \frac{e^2}{V^2} \sum_{kk'} \int_C dt_1 \int_C dt_2 ~
    \mbox{Tr} \left\{  D(t,t') V_L 
     G_{cL} (k'; t', t_2) V_L^{\dag} D(t_2,t_1) V_L 
     G_{cL} (k; t_1, t) V_L^{\dag}  \right\}, \\
\hbar G_{II}^{<} (\epsilon) 
 &=& - \frac{2e^2}{h} \int d\zeta ~ \bar{f}_L (\zeta) 
     \mbox{Tr} \{ D^{<} (\epsilon+\zeta) \Gamma_L \} 
    \nonumber\\
 && - \frac{e^2}{h} \int d\zeta ~ \mbox{Tr} 
   \left\{ D^{<} (\epsilon+\zeta) \Gamma_L
      \left[ D^{>} (\zeta) 
         + 2i \bar{f}_L (\zeta) \{ D^{a} (\zeta) 
             - D^{r} (\zeta) \}
      \right] \Gamma_L
   \right\}.
\end{eqnarray}
Collecting all contributions, the lesser current Green's function is 
given by the expression, 
\begin{eqnarray}
\hbar G_{II}^{<} (\epsilon)
 &=& \frac{e^2}{h} \int d\zeta \mbox{Tr} [ 2f_L(\epsilon+\zeta) 
    D^{r}(\epsilon+\zeta) + i D^{<} (\epsilon+\zeta) ] 
   \Gamma_L [ 2\bar{f}_L(\zeta) D^{r}(\zeta)
     + i D^{>} (\zeta) ] \Gamma_L \nonumber\\
 && + \frac{e^2}{h} \int d\zeta \mbox{Tr} [ 2f_L(\epsilon+\zeta) 
    D^{a}(\epsilon+\zeta) - i D^{<} (\epsilon+\zeta) ] 
   \Gamma_L [ 2\bar{f}_L(\zeta) D^{a}(\zeta)
     - i D^{>} (\zeta) ] \Gamma_L \nonumber\\
 &&  - \frac{2e^2}{h} \int d\zeta ~ f_L (\epsilon+\zeta) 
     \mbox{Tr} \{ \Gamma_L D^{>} (\zeta)  \} 
    \nonumber\\
 && - \frac{e^2}{h} \int d\zeta ~ \mbox{Tr} 
   \left\{ \Gamma_L
      \left[ D^{<} (\epsilon+\zeta) 
         + 2i f_L (\epsilon+\zeta) \{ D^{a} (\epsilon+\zeta) 
             - D^{r} (\epsilon+\zeta) \}
      \right] \Gamma_L D^{>} (\zeta)
   \right\} \nonumber\\
 && - \frac{2e^2}{h} \int d\zeta ~ \bar{f}_L (\zeta) 
     \mbox{Tr} \{ D^{<} (\epsilon+\zeta) \Gamma_L \} 
    \nonumber\\
 && - \frac{e^2}{h} \int d\zeta ~ \mbox{Tr} 
   \left\{ D^{<} (\epsilon+\zeta) \Gamma_L
      \left[ D^{>} (\zeta) 
         + 2i \bar{f}_L (\zeta) \{ D^{a} (\zeta) 
             - D^{r} (\zeta) \}
      \right] \Gamma_L
   \right\}.
\end{eqnarray}
\end{widetext}
The current noise is then given by $S(\epsilon) = - \hbar G_{II}^{<}(\epsilon) 
- \hbar G_{II}^{<}(-\epsilon)$. 
In particular, the $\omega=0$ component current noise $S_0 = -2 \hbar G_{II}^{<}(0)$ 
is given by the equation, 
\begin{eqnarray}
S_0 &=& \frac{8e^2}{h} \int d\epsilon ~ 
   \mbox{Tr} \Gamma_L  D^{<} 
           \Gamma_L  D^{>} \nonumber\\
 && - \frac{8e^2}{h} \int d\epsilon ~ f_L \bar{f}_L
   \mbox{Tr} \Gamma_L  [ D^{r} \Gamma_L D^{r}
        + D^{a} \Gamma_L D^{a} ]
   \nonumber\\
 && - \frac{8e^2}{h} \int d\epsilon ~ f_L
   \mbox{Tr} \Gamma_L D^{>} \Gamma_L \cdot 
      i [ D^{r} - D^{a} ]  \nonumber\\
 && - \frac{8e^2}{h} \int d\epsilon ~ \bar{f}_L 
   \mbox{Tr} \Gamma_L D^{<} \Gamma_L \cdot 
      i [ D^{r} - D^{a} ]  \nonumber\\
 && + \frac{4e^2}{h} \int d\epsilon ~ \left[ 
    f_L \mbox{Tr} \Gamma_L D^{>} 
    + \bar{f}_L \mbox{Tr} \Gamma_L D^{<} \right].
\end{eqnarray}
The above expression of $S_0$ looks highly asymmetrical in the indices 
of left and right leads or $L$ and $R$. 
Substituting the following identity relations
\begin{eqnarray}
D^{<,>} 
 &=& D^{r} \Sigma^{<,>} D^{a}, \\
\Sigma^{<}
 &=& 2f_L \Gamma_L + 2f_R \Gamma_R, \\
\Sigma^{>}
 &=& 2\bar{f}_L \Gamma_L + 2\bar{f}_R \Gamma_R, \\
i[ D^{r} - D^{a} ] 
 &=& 2 D^{r} \Gamma D^{a},
\end{eqnarray}
we simplify the expression of current noise $S_0$ as
\begin{eqnarray}
S_0 &=& \frac{2e^2}{h} \int d\epsilon [f_L\bar{f}_L + f_R\bar{f}_R]
   \mbox{Tr} {\bf T}^2 \nonumber\\
 && + \frac{2e^2}{h} \int d\epsilon [f_L\bar{f}_R + \bar{f}_L f_R]
   \left[ \mbox{Tr} {\bf T} - \mbox{Tr} {\bf T}^2 \right].
\end{eqnarray}
Here ${\bf T} \equiv 4 \Gamma_L D^r \Gamma_R D^a$.

\section{\label{app_det} Properties of determinant $Z_N(\epsilon)$: Eq.~(\ref{det})}
 In this Appendix, we study the properties of the determinant 
$Z_N (\epsilon)$ which is defined by the Eq.~(\ref{det}).  
We can readily find the iterative relation for the determinant when 
$\Gamma_{L,R} = 0$ and $E_i = 0$ for all $i$'s, 
\begin{eqnarray}
Z_{N+2} (\epsilon) 
 &=& \epsilon Z_{N+1}(\epsilon) - |t_{N+1}|^2 Z_N(\epsilon), \\
Z_2(\epsilon) 
 &=& \epsilon^2 - |t_1|^2, \\
Z_3(\epsilon) 
 &=& \epsilon ( \epsilon^2 - |t_1|^2 - |t_2|^2 ). 
\end{eqnarray}
Especially when $t_i = t$ for all $i$'s, the expression of $Z_N(\epsilon)$ 
can be found in an explicit form, 
\begin{eqnarray}
Z_N(\epsilon) 
 &=& \prod_{k=1}^{N} \left[ \epsilon - 2|t| \cos \frac{k\pi}{N+1} \right].
\end{eqnarray}
$Z_N(\epsilon)$ is the type-II Chebyshev polynomial.

 From the recursive relation, we can deduce that $Z_N(\epsilon)$ is invariant 
under the transformation, $t_i \to -t_i$ for any $i$.
We can also show from the recursive relation 
that $Z_N(\epsilon)$ satisfies the parity relation: 
$Z_N(-\epsilon) = (-1)^N Z_N(\epsilon)$.
This parity relation has the important consequence on the particle-hole
symmetry of split atomic energy level structure. 
$Z_{2N}(\epsilon)$ is the even function of $\epsilon$ and 
$Z_{2N}(0) = (-1)^N |t_1 t_3 \cdots t_{2N-1}|^2$. 
$Z_{2N+1}(\epsilon)$ can be written as 
$Z_{2N+1}(\epsilon) = \epsilon Y_{2N+1} (\epsilon)$ where 
$Y_{2N+1} (\epsilon)$ is the even function of $\epsilon$ and 
$Y_{2N+1} (0) \neq 0$. 
When all the atomic energy levels lie at the Fermi energy, 
the hopping integrals between neighboring sites in the atomic wire
lift the degeneracy but does not change the particle-hole symmetry 
of split energy levels. If $\epsilon_0$ is a solution 
of $Z_N(\epsilon)=0$, so is $-\epsilon_0$ because of the parity relation. 
Since $Z_{N}(0) \neq 0$ for even $N$, 
no split levels lie at the Fermi energy. 
Since $Z_N(0) = 0$ for odd $N$, $\epsilon=0$ is a solution of the secular 
equation $Z_N(\epsilon)=0$. Since $Y_{N}(0) \neq 0$ for odd $N$, only 
one level lies at the Fermi energy for the odd-numbered atomic wire.

 Though the particle-hole symmetry of split atomic energy levels is 
shown explicitly, it can be also proved by noting that the tight-binding 
Hamiltonian is invariant under the particle-hole transformation:
$c_i \to (-1)^i c_{i}^{\dag}$. Obviously the Hamiltonian 
$H = \sum_{i} ( t_i c_{i+1}^{\dag} c_i + H.c. )$ remains 
invariant under this particle-hole transformation.

%
%
%
%

\end{document}